\documentclass[aps,nofootinbib,twocolumn,prd,showpacs,class-pre]{revtex4}
\usepackage{setspace} 
\usepackage{amsmath,amssymb,amsfonts,amsthm}
\usepackage{graphicx}

\newcommand{\be}{\begin{equation}}
\newcommand{\ee}{\end{equation}}
\newcommand{\beq}{\begin{eqnarray}}
\newcommand{\eeq}{\end{eqnarray}}
\newcommand{\br}{{\bf r}}
\newcommand{\bR}{{\bf R}}

\newcommand{\p}{\partial}
\newcommand{\half}{\frac{1}{2}}
\newcommand{\nn}{\nonumber}

\def\lsim{\hbox{ \raise.35ex\rlap{$<$}\lower.6ex\hbox{$\sim$}\ }}
\def\gsim{\hbox{ \raise.35ex\rlap{$>$}\lower.6ex\hbox{$\sim$}\ }}

\begin{document}
\title{Gravitational Waves in the Spectral Action of Noncommutative Geometry}
\author{William Nelson\footnote{nelson@gravity.psu.edu}, Joseph
  Ochoa\footnote{jro166@psu.edu}} \affiliation{ Institute of
  Gravitation and the Cosmos, Penn State University, State College, PA
  16801, U.S.A.}  \author{Mairi
  Sakellariadou\footnote{mairi.sakellariadou@kcl.ac.uk}}
\affiliation{Department of Physics, King's College, University of
  London, Strand WC2R 2LS, London, U.K.}

\begin{abstract}
The spectral triple approach to noncommutative geometry allows one to
develop the entire standard model (and supersymmetric extensions) of
particle physics from a purely geometry stand point and thus treats
both gravity and particle physics on the same footing. The bosonic
sector of the theory contains a modification to Einstein-Hilbert
gravity, involving a nonconformal coupling of curvature to the Higgs
field and conformal Weyl term (in addition to a nondynamical
topological term). In this paper we derive the weak field limit of
this gravitational theory and show that the production and dynamics of
gravitational waves are significantly altered. In particular, we show
that the graviton contains a massive mode that alters the energy lost
to gravitational radiation, in systems with evolving quadrupole
moment. We explicitly calculate the general solution and apply it to
systems with periodically varying quadrupole moments, focusing in
particular on the the well know energy loss formula for circular
binaries.
\end{abstract}


\maketitle

\section{Introduction}
\label{intro}
Noncommutative Geometry (NCG) is a gravitational theory which, even in
its simplest form, can explain the Standard Model of particle physics,
and account for all current experimental data, in a rather simple and
certainly elegant way.  The simple --- in the sense that it
generalizes the continuum Riemaniann manifold by considering its
product by a discrete two points space --- NCG proposal should be
certainly replaced by a less trivial noncommutative space as one
reaches Planckian energy scales. Nevertheless, this is the proposal we
have at hand, and given its success in accounting for the Standard
Model of particle physics, it offers a valid theoretical framework to
address early universe issues. Alternatively, one can use
experimental high energy physics data and astrophysical
observations/measurements in order to test this NCG proposal and
constrain its parameters. This is the approach used in this study.

One should indeed view this NCG proposal as an effective theory, which
can however offer a valuable information about any NCG approach. In
what follows we derive explicitly the weak field limit of this
gravitational theory and then show that the production and dynamics of
gravitational waves are both considerably modified from those obtained
within the familiar General Relativity approach.

More precisely, in Section~\ref{NCG-section} we give a short
introduction to the noncommutative geometry spectral action, the
framework within which we will then focus. In
Section~\ref{pert-equations} we first state in detail the conventions and
signature we use and we analyze the issue of gauge conditions. We then
analyze linear perturbations around a Minkowski background metric and
we solve the noncommutative geometry gravitational wave equation in
terms of the retarded Green's function.  We find that gravitational
waves are only sourced from systems with a nontrivial quadrupole
moment, as within General Relativity, while the NCG theory contains
massive as well as massless gravitons. In Section~\ref{examples} we
concentrate first on some simple and then on some physical
examples. Using the requirement that the mass of the gravitons must be
positive (and real), we can fix the sign of the couplings in the NCG
spectral action approach.  We then calculate the energy loss for a
circular binary system and compare it to the results obtained from
standard General Relativity. We conclude that the amplitude of
modifications within NCG is small, nevertheless the NCG approach leads
to some distinctive features which we analyze.  We round up with our
conclusions in Section~\ref{conclusions}.

\section{Noncommutative Geometry Spectral Action}
\label{NCG-section}
In the NonCommutative Geometry~\cite{ncg-book1,ncg-book2} approach,
the Standard Model (SM) of electroweak and strong interactions is
considered as a phenomenological model, which dictates the geometry of
space-time, so that the associated Maxwell-Dirac action functional
produces the SM with all known experimental results.  The outcome of
this approach is a geometric space defined by the product, ${\cal
  M}\times{\cal F}$, of a continuum compact Riemaniann manifold,
${\cal M}$, and a tiny discrete finite noncommutative space, ${\cal
  F}$, composed of only two points.  Such an almost commutative space
is the simplest extension of the more familiar commutative space upon
which General Relativity is formulated. Certainly one should not
expect the validity of this simplistic approach to hold at the Planck
scale, which is the scale at which all notion of classical geometry loses its
meaning.

The metric dimension of the product geometry ${\cal M}\times{\cal F}$
is 4, the same as the ordinary space-time manifold. Thus, the metric
dimension of the noncommutative space ${\cal F}$ is zero, while
for noncommutative spaces one must distinguish between the metric
dimension and the $KO$-dimension. The internal space ${\cal F}$ has
$KO$-dimension $6$ to allow fermions to be simultaneously Weyl and
chiral, whilst it is discrete to avoid the infinite tower of massive
particles that are produced in string theory.  

The noncommutative nature of ${\cal F}$ is given by the real spectral
triple $({\cal A}, {\cal H}, D)$ that generalizes Riemannian geometry
to the noncommutative setting; ${\cal A}$ is an involution of
operators on the finite-dimensional Hilbert space ${\cal H}$ of
Euclidean fermions, and $D$ is a self-adjoint unbounded operator in
${\cal H}$.  The choice of Hilbert space has no importance, since all
separable infinite-dimensional Hilbert spaces are isomorphic.  The
algebra ${\cal A}$, related to the gauge group of local gauge
transformations, is the algebra of coordinates. A space is described
by the algebra of coordinates, which in the context of NCG is
represented as operators on a Hilbert space. Since real coordinates
are represented by self-adjoint operators, all information about a
space within NCG is encoded in the algebra of coordinates ${\cal A}$.
By assuming that the algebra constructed in ${\cal M}\times {\cal F}$
is symplectic-unitary, ${\cal A}$ must be of the form
\begin{equation}
\mathcal{A}=M_{a}(\mathbb{H})\oplus M_{k}(\mathbb{C})~;
\end{equation}
$k=2a$, $\mathbb{H}$ is the algebra of quaternions.  The choice $k=4$
is the first value that produces the correct number of fermions in
each generation, {\sl i.e.}, $k^2=16$ fermions in each of the three
generations~\cite{Chamseddine:2007ia}.  

The operator $D$ corresponds to the inverse of the Euclidean
propagator of fermions, and is given by the Yukawa coupling matrix
which encodes the masses of the elementary fermions and the
Kobayashi--Maskawa mixing parameters. The commutator $[D,a]$, with
$a\in {\cal A}$, plays the r\^ole of the differential quotient
$da/ds$, with $ds$ the unit of length.  The familiar geodesic formula
\be
d(x,y)={\rm inf}\int_\gamma ds~,
\ee
where the infimum is taken over all possible paths connecting $x$ to
$y$, which is used to determine the distance $d(x,y)$ between two
points $x$ and $y$ within Riemannian geometry, is replaced by
\be
d(x,y)={\rm sup}\{|f(x)-f(y)|: f\in {\cal A}, ||[D,f]|| \leq 1\}~,
\ee
where $D$ is the inverse of the line element $ds$, within the
noncommutative spectral geometry.

The fermions of the SM provide the Hilbert space ${\cal H}$ of a
spectral triple for the algebra ${\cal A}$, while the bosons of the
SM, including the Higgs boson, are obtained through inner fluctuations
of the Dirac operator of the product ${\cal M}\times {\cal F}$
geometry. Hence, the Higgs boson, which generates the masses of
elementary particles through spontaneous symmetry breaking, becomes
just a gauge field corresponding to a finite difference.  Note that
the corresponding mass scale specifies the inverse size of the
discrete geometry ${\cal F}$.

Applying the spectral action principle, according to which the action
functional on spectral triples depends only on the spectrum of the
line element, {\sl i.e.}, the inverse of the Dirac operator, to the inner
fluctuations of the product geometry ${\cal M}\times {\cal F}$, one
recovers the SM coupled to gravity in the Euclidean form.  Thus, the
NCG spectral action approach --- limited to the classical level even
though it can {\sl a priori} be quantized --- offers an elegant
geometric interpretation of the SM , the most successful
phenomenological model of particle physics.

To be more precise, the SM Lagrangian --- including mixing and
Majorana mass terms for neutrinos, minimally coupled to gravity ---
can be successfully recovered from the asymptotic expansion of the
spectral action functional
\be\label{spectral-action}
{\rm  Tr}\Big(f\Big(\frac{D}{\Lambda}\Big)\Big)~,\ee
where $f$ is a positive even function of the real variable and
$\Lambda$ fixes the energy scale. Note that $D/\Lambda$ is
dimensionless since the Dirac operator, being a differential operator,
has dimensions of mass.  The physical Lagrangian is thus obtained from
the asymptotic expansion in the energy scale $\Lambda$ of the spectral
action functional, Eq.~(\ref{spectral-action}).  More precisely, using
heat kernel methods one can write the square of the Dirac operator in
terms of the inverse metric, the unit matrix and two matrix functions
computed from $D$ and show that the trace,
Eq.~(\ref{spectral-action}) above, can be expanded in a power series
as a function of the inverse scale $\Lambda$ and it can thus be
written in terms of the geometrical Seeley-deWitt coefficients $a_n$,
as~\cite{sdw-coeff}
\be
\sum_{n=0}^\infty F_{4-n}\Lambda^{4-n}a_n~,
\ee
where the function $F$ is defined such that $F(D^2)=f(D)$.
Defining the moments
\be\label{eq:moments0}
f_k=\int_0^\infty f(u) u^{k-1}{\rm d}u\ \ ,\ \ \mbox{for}\ \ k>0 ~,
\ee
and $f_0=f(0)$, one finds
\beq \label{eq:moments}
 F_4&=&2f_4\nonumber\\ F_2&=&2f_2\nonumber\\ F_0&=&f_0\nonumber\\
F_{-2n}&=&\Big[(-1)^n\Big(\frac {\rm d}{2u{\rm d}u}\Big)^n
  f\Big](0)\ \ \mbox{for}\ \ n\geq 1~,
\eeq
while the coefficients $a_n$ are known for any second order elliptic
differential operator.

The coupling with fermions can be obtained by including an additional
fermionic term
\be
\frac{1}{2}<J\psi,D\psi>~,
\ee
in Eq.~(\ref{spectral-action}), where $J$ is the real structure on the
spectral triple and $\psi$ is a spinor in the Hilbert space ${\cal H}$
of the quarks and leptons.

The spectral action approach leads naturally to the merging of the
three coupling constants at the unification scale, $g_2=g_3=\sqrt{5/3}
g_1$, it provides neutrino masses and mixing as well as the see-saw
mechanism, and it predicts a heavy Higgs mass.

The spectral action, Eq.~(\ref{spectral-action}), can be expanded in
powers of the scale $\Lambda$ in the form
\begin{equation}
{\rm Tr}\left(f\left(\frac{D}{\Lambda}\right)\right)\sim 
\sum_{k\in {\rm DimSp}} f_{k} 
\Lambda^k{\int\!\!\!\!\!\!-} |D|^{-k} + f(0) \zeta_D(0)+ {\cal O}(1)~,
\end{equation}
where $f_k$ are the momenta of the function $f$ given in
Eq.~(\ref{eq:moments0}), the noncommutative integration is defined in
terms of residues of zeta functions, and the sum is over points in the
{\sl dimension spectrum} of the spectral triple.

The physical Lagrangian that one obtains in this approach, contains,
in addition to the full SM Lagrangian, the Einstein-Hilbert action
with a cosmological term, a topological term related to the Euler
characteristic of the space-time manifold, a conformal Weyl term and a
conformal coupling of the Higgs field to gravity. Note that the
coefficients of the gravitational terms depend on the Yukawa
parameters of the particle physics content. Within the NCG spectral
action, one works in Euclidean rather than Lorentzian signature,
assuming that one can get back to the Minkowski signature through 
Wick rotation.

One then sets the parameters of the NCG spectral action at the
(unification) scale $\Lambda$; predictions at lower energies are
recovered by running the parameters down through Renormalization Group
Equations (RGE). Hence, the spectral action at the unification scale
$\Lambda$ offers a framework to investigate early universe
cosmological
models~\cite{Nelson:2008uy,Nelson:2009wr,Marcolli:2009in,mmm,wjm}, while
extrapolations to lower energies can be obtained via, firstly, RGE and
secondly, inclusion of nonperturbative effects in the spectral action.

Adopting Euclidean signature, the gravitational part of the asymptotic
formula for the bosonic sector of the NCG spectral action, including
the coupling between the Higgs field $\phi$ and the Ricci curvature
scalar $R$, is~\cite{ccm}
\beq\label{eq:0}{\cal S}_{\rm grav}^{\rm E} = \int \left(
\frac{1}{16\pi G} R + \alpha_0
C_{\mu\nu\rho\sigma}C^{\mu\nu\rho\sigma} + \tau_0 R^\star
R^\star\right.  \nonumber\\ -\left.  \xi_0 R|{\bf H}|^2 \right)
\sqrt{g} {\rm d}^4 x~. \eeq
Note that ${\bf H}$ is a rescaling ${\bf H}=(\sqrt{af_0}/\pi)\phi$ of
the Higgs field $\phi$ to normalize the kinetic energy; the momentum
$f_0$ is physically related to the coupling constants at unification
and the coefficient $a$ is related to the fermion and lepton masses
and lepton mixing.  

In the above action, Eq.~(\ref{eq:0}), the first two terms only depend
upon the Riemann curvature tensor; the first is the Einstein-Hilbert
term with the second one being the Weyl curvature term. The third term
\be R^\star
R^\star=\frac{1}{4}\epsilon^{\mu\nu\rho\sigma}\epsilon_{\alpha\beta\gamma\delta}
R^{\alpha\beta}_{\mu\nu}R^{\gamma\delta}_{\rho\sigma}~,\nonumber\ee
is the topological term that integrates to the Euler characteristic
and hence is nondynamical.  The spectral action contains one more term
that couples gravity with the SM, namely the last term in
Eq.~(\ref{eq:0}), which should always be present when one considers
gravity coupled to scalar fields. This coupling can have significant
consequences at high energies, such as in the early
universe~\cite{Nelson:2008uy,Nelson:2009wr,Marcolli:2009in,mmm,wjm},
however in this paper we will be concerned with the low energy weak
curvature regime where this term is small.

Neglecting the nonminimal coupling between the Higgs field and the
Ricci curvature, the equations of motion derived from the Lorentzian
version of spectral action above read~\cite{Nelson:2008uy}
\beq\label{eq:EoM2} R^{\mu\nu} - \frac{1}{2}g^{\mu\nu} R -
 32\pi G\alpha_0 \left[ 2C^{\mu\lambda\nu\kappa}_{;\lambda ; \kappa}
  + C^{\mu\lambda\nu\kappa}R_{\lambda \kappa}\right]\nonumber\\ =   
\ 8\pi G T^{\mu\nu}_{\rm matter}~,  \eeq                            
implying that the NCG corrections vanish~\cite{Nelson:2008uy} for
Friedmann-Lema\^{i}tre-Robertson-Walker (FLRW) cosmologies. [The
  reader is directed to subsection~\ref{pert-equations}A for a definition
  and discussion of the Lorentzian conventions used.]

We will be concerned with linear perturbations around a Minkowski
background metric in the synchronous gauge, so that the perturbed
metric reads
\be g_{\mu\nu} = {\rm diag} \left( \{a(t)\}^2 \left[
  -1,\left(\delta_{ij} + h_{ij}\left(x\right)\right) \right]\right)~,
\ee
where $a(t)$ is the cosmological scale factor. Throughout this paper
we work in a flat background and hence $a(t)=1$ and $\dot a\equiv{\rm
  d}t/{\rm d}t=0$.  The remaining gauge freedom can be completely
fixed by setting ${\bf \nabla}_i h^{ij}=0$. [A detailed discussion of
this gauge fixing is given in subsection~\ref{pert-equations}C below.]

In Section~\ref{pert-equations} below we show that, the linearized
equations of motion derived from the NCG spectral action,
for such perturbations, read 
\be\label{eq:1} \left( \Box - \beta^2 \right) \Box h^{\mu\nu} =
\beta^2 \frac{16\pi G}{c^4} T^{\mu\nu}_{\rm matter}~, \ee 
where $T^{\mu\nu}_{\rm matter}$ is taken to lowest order in
$h^{\mu\nu}$. This implies that it is independent of $h^{\mu\nu}$
and satisfies the conservation equations
\be
\frac{\partial}{\partial x^\mu} T^\mu_{\ \nu}=0~.
\ee
It is important to note that $\beta$, defined as
\be \beta^2 \equiv -1/(32\pi G\alpha_0)~,
\ee
in Eq.~(\ref{eq:1}), turns out to play the r\^ole of a mass and hence
has to be real and positive, implying that $\alpha_0 <0$.  In the
following we will see that, for $\alpha_0>0$, the gravitational waves
evolve according to a Klein-Gordon like equation with a tachyonic
mass, and hence the background, which in our case is Minkowski space,
is unstable.  We can thus restrict to $\alpha_0<0$ for Minkowski space
to be a (stable) vacuum of the theory.

\section{Perturbation Equations}~\label{pert-equations}
To write down the linearized equations of motion, we will first discuss
our conventions for the metric signature and the Ricci tensor.

\subsection{Conventions for $R_{\mu\nu}$ and Signature}~\label{conventions}
In this paper, we are using conventions in which the signature is
$(-,+,+,+)$ and the Ricci tensor is defined as $R_{\mu\nu} = 
R^\rho\phantom{}_{\mu\nu\rho}$,
with $R_{\mu\nu\rho}\phantom{}^\sigma\omega_\sigma =
 \big[ \bigtriangledown_\mu , \bigtriangledown_\nu \big] \omega_\rho$. 
In General Relativity such choices
are merely conventions, which are relatively unimportant (provided of
course that one is consistent), here however the situation is very
different. The Lorentzian version of the NCG action, Eq.~(\ref{eq:0}),
that we use reads
\beq\label{eq:1.5} {\cal S}_{\rm grav}^{\rm L} = \int \left(
\frac{1}{16\pi G} R + \alpha_0
C_{\mu\nu\rho\sigma}C^{\mu\nu\rho\sigma} + \tau_0 R^\star
R^\star\right.  \nonumber\\ -\left.  \xi_0 R|{\bf H}|^2 \right)
\sqrt{-g} {\rm d}^4 x~. \eeq
It is thus clear that the conventions used to define, for example,
$R_{\mu\nu}$ will radically alter the theory, unless one also alters
the (signs) of the couplings. Specifically, consider using the
opposite convention for the Ricci tensor, which introduces a negative
sign on all terms depending on $R_{\mu\nu}$, but {\it not} on terms
depending on $R_{\mu\nu\rho\sigma}$. Since our action now contains
terms of both kinds ({\sl i.e.}, $R$ and
$C_{\mu\nu\rho\sigma}C^{\mu\nu\rho\sigma}$), this change of convention
introduces a relative sign change. This can simply be compensated for
by changing $\alpha_0 \rightarrow -\alpha_0$. However, without this
change the action is very different.  An exactly similar change
happens if we considered a different choice of convention for the
signature or the sign of $R_{\mu\nu\rho\sigma}$, as these both
introduce a sign change in the $R$ term, but not the
$C_{\mu\nu\rho\sigma}C^{\mu\nu\rho\sigma}$ term.

As shown in
Refs.~\cite{Nelson:2008uy,Nelson:2009wr,Marcolli:2009in,mmm}, the
presence of the non-minimal coupling of curvature to the Higgs field
can have significant effects of the cosmological dynamics and one may
wonder whether the sign ambiguities discussed here may affect these
results. Fortunately, from Eq.~(\ref{eq:1.5}) it is clear that the
relative sign between the Einstein-Hilbert term ($R$) and the
nonminimal coupling ($R|{\bf H}|^2$) is independent of any convention
(since they both contain $R$). In a cosmological setting ({\sl i.e.},
for FLRW geometries) the Weyl term vanishes and hence, in homogeneous
cosmologies, the only NCG affects come from the nonminimal
coupling~\cite{Nelson:2008uy}, allowing such issues to be avoided.

Since the underlying NCG theory is only developed for the Euclidean
signature, it does not provide a guide for the Wick rotation to the
Lorentzian space. Hence the choice of the sign of the couplings,
appropriate for a particular choice of convention, can only be made by
testing the {\it physical} consequences of the theory. In the
following section we will show that gravitational waves offer an
excellent probe of the couplings in this theory, but even without an
in-depth analysis, from Eq.~(\ref{eq:1}) one can immediately see that
the coupling $\alpha_0$ must be negative (in the conventions used
here). If it were not, the $\beta$ parameter would be complex and this
would correspond to a tachyonic mode of the graviton (we refer the
reader to a discussion below).  This would indicate that Minkowski
space-time is unstable to small perturbations. If we rule out such a
possibility on physical grounds (or require that the Lorentzian
version of the NCG action should admit Minkowski as a stable vacuum)
then we can restrict $\alpha_0<0$ {\it with the conventions used
  here}. Consider for example the consequence of changing the
signature, so that $\Box \rightarrow -\Box$. Such a change of
convention essentially changes the sign of $\beta^2$ and the
conclusions would be reversed\footnote{The concerned reader should
  note that exactly the same situation arises in standard Klein-Gordon
  equation, where the sign of the mass term is changed under a change
  of signature.}.

\subsection{Linearized Equations of Motion}\label{linear-pert}

Variation of the gravitational part action Eq.~(\ref{eq:1.5}) {\sl
  w.r.t.} the metric $g_{\mu\nu}$ leads to the following addition to
the Einstein tensor of General Relativity (GR),
\beq
\label{LEOM:1}
 G\phantom{}_{\text{NCG}}^{\mu\nu}&=&
 -\frac{1}{2\kappa}G_\text{Einstein}^{\mu\nu} \\ \nn &&\, +2\alpha_0\bigg
 (2\bigtriangledown_\lambda \bigtriangledown_\kappa
 C^{\mu\kappa\nu\lambda} + C^{\mu\kappa\nu\lambda}
 R_{\kappa\lambda}\bigg) \, , \eeq 
where as usual
$$G_\text{Einstein}^{\mu\nu} = R^{\mu\nu} - \frac{1}{2}g^{\mu\nu}
R~.$$
Given the convention used here to define the Ricci tensor the Weyl
Tensor is explicitly given as, 
\beq
\label{LEOM:2}
 C_{\mu\lambda\nu\kappa} &=& R_{\mu\lambda\nu\kappa} + (g_{\mu[\nu}R_{\kappa]
\lambda}-g_{\lambda[\nu}R_{\kappa]\mu})\\
\nn &&\qquad-\frac{1}{3}g_{\mu[\nu}g_{\kappa]\lambda}R \, .
\eeq
Using the contracted Bianchi identity
\beq
\label{LEOM:3}
\bigtriangledown^\kappa R_{\mu\lambda\nu\kappa} = -
(\bigtriangledown_\lambda R_{\mu\nu} - \bigtriangledown_\mu
R_{\lambda\nu} )\, , \eeq
and its remaining trace
\beq
\label{LEOM:4}
 \bigtriangledown^\kappa R_{\lambda\kappa} &=&
 \half\bigtriangledown_\lambda R \, , \eeq 
we can arrive at the following expression 
\beq
\label{LEOM:5}
2\bigtriangledown^\lambda\bigtriangledown^\kappa
C_{\mu\lambda\nu\kappa} &=& - C_{\lambda\mu\kappa\nu}R^{\lambda\kappa}
-\bigtriangledown^\lambda\bigtriangledown_\lambda\big( R_{\mu\nu} -
\frac{1}{6}g_{\mu\nu} R \big) \nonumber\\ &&\,\,+
\frac{1}{3}\bigtriangledown_\mu \bigtriangledown_\nu R - 2R_{\mu\rho}
R^\rho\phantom{}_{\nu} + \frac{2}{3}RR_{\nu\mu} \nonumber \\ 
&&
\,\,\, + \frac{1}{2}g_{\mu\nu}\big( R_{\kappa
  \lambda}R^{\lambda\kappa}-\frac{1}{3}R^2\big) \,~.  \eeq
Notice that the expression above for $2\bigtriangledown_\lambda
\bigtriangledown_\kappa C^{\mu\lambda\nu\kappa}$ shows that the
$C^{\mu\kappa \nu\lambda} R_{\kappa\lambda}$ term in
Eq.~(\ref{LEOM:1}) exactly cancels in favor of terms which are of
second order of solely the Ricci tensor and/or Ricci scalar.

We now follow the standard procedure of perturbing about a flat
metric, where 
\be
g_{\mu\nu} = \eta_{\mu\nu} + \gamma_{\mu\nu}~~,~~
g^{\mu\nu} = \eta^{\mu\nu} - \gamma^{\mu\nu}~,
\ee
and
\be
\gamma =
\gamma^\mu\phantom{}_\mu = \eta^{\mu\nu}\gamma_{\mu\nu}~;
\ee
all tensor indices are raised and lowered using the background metric
$\eta_{\mu\nu}$ (except for the indices of $g_{\mu\nu}$ and
$g^{\mu\nu}$). To first order in metric perturbations we then have
\beq
\label{LEOM:6}
&& 2\Big\{\bigtriangledown _\lambda\bigtriangledown _\kappa C^{\mu
  \kappa \nu \lambda}\Big\} =\nonumber \\ && \p_\lambda\p^\lambda\Big
( \p^\kappa\p^{(\nu}\bar\gamma^{\mu)}\phantom{}_\kappa -
\half\p_\kappa\p^\kappa\bar\gamma^{\mu\nu} -\frac{1}{6}\eta^{\mu\nu}
\p^\sigma\p^\kappa\bar\gamma_{\sigma\kappa} \Big )\nonumber\\ && \,\,
-\p_\lambda\p^\nu\Big ( \p^\kappa\p^{(\lambda}\bar\gamma^{\mu)}
\phantom{}_\kappa - \half\p_\kappa\p^\kappa\bar\gamma^{\lambda\mu}
\Big )
+\frac{1}{6}\p^\mu\p^\nu\p^\lambda\p^\kappa\bar\gamma_{\lambda\kappa}\nonumber
\\ &&\,\,\,\, -\frac{1}{6}\Big (\eta^{\mu\nu}\p_\kappa\p^\kappa -
\p^\mu \p^\nu\Big )\p_\lambda\p^\lambda\gamma +\mathcal{O}(\gamma^2
)\, .  \eeq
where $\mathcal{O}(\gamma^2 )$ denotes any second order combinations of 
$\gamma_{\mu\nu}$ and we have defined  
\be
\label{LEOM:7}
\bar\gamma_{\mu\nu} = \gamma_{\mu\nu} - \half\eta_{\mu\nu}\gamma \, ,
\ee 
{\sl i.e.}, the {\sl trace reverse} of $\gamma_{\mu\nu}$. 

Similarly, to linear order in metric perturbations the Einstein tensor
is simply
\be
\label{LEOM:8}
{G}\phantom{}_\text{Einstein}^{\mu\nu}= + \half \p_\lambda\p^\lambda
\bar\gamma^{\mu\nu} +\mathcal{O}(\gamma^2 )~.  \ee
%

\subsection{Gauge Conditions}\label{gauge-subsection}
In calculating~\cite{Nelson:2008uy} the linearized equations of motion,
the traceless transverse gauge was imposed on the metric perturbations
$h_{\mu\nu}$; here we explicitly show that this is indeed a valid
choice. As before, we denote metric perturbations that have not been
gauge fixed by $\gamma_{\mu\nu}$ and reserve $h_{\mu\nu}$ for the
final, gauge fixed perturbations that correspond to the physical
gravitational waves.

As always we have a freedom due to diffeomorphism invariance of the
action to restrict the gauge of the metric perturbations. Explicitly,
under a diffeomorphism generated by $\xi_\mu$ the metric perturbations
$\gamma_{\mu\nu}$ transform as 
\be
\label{gc:3}
\gamma_{\mu\nu}^{\rm old}\stackrel{\xi_\mu}{\longrightarrow}
\gamma^{\rm new}_{\mu\nu} = \gamma_{\mu\nu}^{\rm old} + \p_\mu \xi_\nu
+\p_\nu\xi_\mu \, .  \ee
Without loss of generality, one can impose the {\sl Lorentz gauge
  conditions}
\be
\label{gc:5}
\p^\mu\bar\gamma_{\mu\nu} = 0 \, ,
\ee
restricting the perturbations to be transverse, where we introduced
the ``trace reverse'' of $\gamma_{\mu\nu}$, as in Eq.~(\ref{LEOM:7}).

Choosing this gauge (and dropping the label {\sl new} on
$\gamma_{\mu\nu}$), Eq.~(\ref{LEOM:6}) simplifies to
\beq
\label{gc:6}
 \nn 2\Big\{\bigtriangledown _\lambda\bigtriangledown _\kappa
C^{\mu \lambda\nu\kappa}\Big\} = \nonumber \\ -\half
\p_\kappa\p^\kappa \Big (\p_\lambda\p^\lambda\bar\gamma^{\mu\nu}
+\frac{1}{3}\big (\eta^{\mu\nu}\p_\lambda\p^\lambda - \p^\mu\p^\nu\big
)\gamma \Big)\, .  \eeq
Combining the above result with the Einstein contribution, $G_{\rm
  Einstein}^{\mu\nu}$, to the equation of motion, the left-hand side
of Eq.~(\ref{eq:EoM2}) is, to first order in $\gamma_{\mu\nu}$, given
by
\beq
\label{gc:7}
 \nn {G}\phantom{}_{\text{NCG}}^{\mu\nu} =
     -{G}\phantom{}_\text{Einstein}^{\mu\nu}
     -\frac{1}{\beta^2}\Big\{2\bigtriangledown
     _\kappa\bigtriangledown _\lambda C^{\mu \lambda\nu\kappa} \Big\}
     \, \ \ \ \ \ \ \ \ \nonumber\\
= -\half
     \p_\kappa\p^\kappa\bar\gamma^{\mu\nu}\ \ \ \ \ \ \ \ \ \ \ \ \ 
\ \ \ \ \ \ \ \ \ \ \ \ \ \ \ \nonumber
     \\ +\frac{1}{2\beta^2} \p_\kappa\p^\kappa\Big
     (\p_\lambda\p^\lambda \bar\gamma^{\mu\nu} +\frac{1}{3}\big
     (\eta^{\mu\nu}\p_\lambda\p^\lambda - \p^\mu\p^\nu\big )\gamma
     \Big)\, .  \eeq
However, the Lorentz gauge does not uniquely fix all the gauge
freedom. More precisely, we still are free to perform gauge
transformations, generated by any $\xi_\mu$, that satisfy
\be
\label{gc:8}
\p_\mu\p^\mu\xi_\nu = 0~,
\ee
since this still preserves the gauge condition, Eq.~(\ref{gc:5}), as
can be checked directly. 

We can use this transformation, Eq.~(\ref{gc:8}), to set (in the new
frame) $\gamma = 0$ and $\gamma_{0i} =0 \,\, (i=1,2,3)$ by solving the
corresponding equations for $\xi_i$ and their time derivatives on some
initial surface $t=t_0$ where no sources are present and further
extending into a source free region ({\sl i.e.},
$T^{\mu\nu}=0$)~\cite{R_Wald}.  After performing these gauge
transformations the source free equations of motion ({\sl i.e.}, the
left-hand side of Eq.~(\ref{eq:EoM2})) read
\beq
\label{gc:11}
-\half \p_\kappa\p^\kappa\bar\gamma_{\mu\nu}
+\frac{1}{2\beta^2}
\p_\kappa\p^\kappa\p_\lambda\p^\lambda\bar \gamma_{\mu\nu} &=& 0\,\,
, \eeq
where we have made repeated use of the fact that
$\p_\mu\p^\mu\xi_\nu=0$.

Up to this point, the gauge restrictions $\gamma = 0$ and
$\gamma_{0i}=0 \,\, (i=1,2,3)$ are the same as those typically used in
General Relativity, which is to be expected since all we have used is
the diffeomorphism invariance of the action.  However, in determining
whether we can set $\gamma_{00} =0$, the equations of motion are used
and hence one might expect that this gauge condition will be different
than that of General Relativity.  To confirm its validity note that
Eq.~(\ref{gc:5}) implies
\be
\label{gc:12}
\frac{\p \gamma_{00}}{\p t} = 0 \, .  \ee

Now, using the equation of motion in the presence of matter source we
arrive at
\beq
\label{gc:13}
\nn \boldsymbol{\bigtriangledown}^2\gamma_{00}- \frac{1}{\beta^2}
\boldsymbol{\bigtriangledown}^2\big(\boldsymbol{\bigtriangledown}^2
\gamma_{00}\big) &=&-\frac{2\kappa}{c^4} T_{00} \, \, ,\\ \Big(1-
\frac{1}{\beta^2} \boldsymbol{\bigtriangledown}^2\Big)
\boldsymbol{\bigtriangledown}^2\gamma_{00} &=&-\frac{2\kappa}{c^4}
T_{00}\, .  \eeq
Recall that General Relativity is recovered in this setting by taking
$\beta\rightarrow \infty$. Thus, one can see that in this limit the
equation simplifies to $\boldsymbol{\bigtriangledown}^2\gamma_{00}
=-(16\pi G)/c^4 T_{00}$, which fixes $\gamma_{00}$ to be a constant
(assuming the space-time is asymptotically flat) away from the
source. Finally, a redefinition (gauge transformation) allows us to
set $\gamma_{00}=0$.

From Eq.~(\ref{gc:13}) we see that, away from the source,
$\boldsymbol{\bigtriangledown}^2\gamma_{00}=0$ is still a solution and
hence we can fix $\gamma_{00}=0$, however this is no longer the only
solution to Eq.~(\ref{gc:13}). In particular, away from the source one
could fix $\gamma_{00}$ via,
\be\label{eq:mod_rad_gauge} \Big(1 - \frac{1}{\beta^2}
\boldsymbol{\bigtriangledown}^2\Big) \gamma_{00}=0~, \ee
which clearly solves Eq.~(\ref{gc:13}). This would result in a
modification of what is often referred to as the {\sl radiation
  gauge}. In the following, we choose $\gamma_{00}=0$ so as to be able
to directly compare our results to the standard ones obtained within
General Relativity.

In cases where sources are present, the NCG equation of motion with
gravity and normal matter is 
\beq
\label{gc:14}
&&
\p_\kappa\p^\kappa\bar\gamma^{\mu\nu}
-\frac{1}{\beta^2} \p_\kappa\p^\kappa\Big
[\p_\lambda\p^\lambda\bar \gamma^{\mu\nu} 
+\frac{1}{3}\big (\eta^{\mu\nu}\p_\lambda\p^\lambda - \p^\mu\p^\nu\big
)\gamma \Big]\, \nonumber\\
&&= -\frac{2\kappa}{c^4} T^{\mu\nu}~.  \eeq
Since $T_{\mu\nu}\neq 0$ we are not free to impose the traceless
condition of the radiation gauge~\footnote{Nor would this be possible
  if one had chosen the modified radiation gauge implied by
  Eq.~(\ref{eq:mod_rad_gauge}).}.  But the explicit presence of the
trace $\gamma$ in Eq.~(\ref{gc:14}) above can be eliminated by
formally defining the tensor $\bar h_{\mu\nu}$ as, 
\be
\label{gc:15}
\bar h_{\mu\nu} = \bar\gamma_{\mu\nu}
-\frac{1}{3\beta^2}\mathcal{O}^{-1}\big (\eta_{\mu\nu} \Box -
\p_\mu\p_\nu\big )\gamma \, , \ee
where the operator $\mathcal{O}$ is given by
\be
\label{gc:16}
\mathcal{O} =\Big(1-\frac{\Box}{\beta^2}\Big) \, \, .
\ee
This is a modification of the {\sl trace reverse} of the metric
perturbations that is usually used, however it performs the same task,
namely removing the trace from the equations of motion,
Eq.~(\ref{gc:14}).  As it can be easily checked, the Lorentz gauge
condition, Eq.~(\ref{gc:5}), is satisfied by $\bar h_{\mu\nu}$ as long
as it is also satisfied by $\bar \gamma_{\mu\nu}$.  Note that the
trace of $\bar h_{\mu\nu}$ is
\be
\label{gc:17}
\bar h=-\Big(1+\frac{\mathcal{O}^{-1}\Box}{\beta^2}\Big)\gamma \,\, ,
\ee
so we see that indeed, this reproduces the {\sl trace reverse} of
$\gamma_{\mu\nu}$ in the limit $\beta\rightarrow \infty$. Clearly
then, when we are away from a source, we can impose that $\gamma=0$
and this implies that $\bar h =0$.

In terms of $\bar h_{\mu\nu}$ the equation of motion,
Eq.~(\ref{gc:14}), is
\beq
\label{gc:18}
\Big (1-\frac{1}{\beta^2}\Box\Big)\Box\bar h^{\mu\nu} \, &=& -\frac{16
  \pi G}{c^4} T^{\mu\nu} \,\, .  \eeq
Dropping the over-bars, this is exactly Eq.~(\ref{eq:1}).


\subsection{Green's Function}\label{sec:grav-wav}
The general physical solution to Eq.~(\ref{gc:18}) is given by
\beq \label{eq:field} h^{\mu\nu} = 2\beta^2\kappa\int {\rm d}S(x')
G_{\rm R}(x,x') T^{\mu\nu}(x')\, , \eeq
where the Green's function $G_{\rm R}(x,x')$ satisfies the 
fourth-order partial differential equation:
\beq
\label{gf:1}
\Big (\Box-\beta^2\Big)\Box G_{\rm R}(x,x') \, &=& 4\pi\delta^{(4)}(x-x') \, ,
\eeq
where the operators $\Box$ above are acting on $x$. In order to find a
solution $G_{\rm R}(x,x')$ to Eq.~(\ref{gf:1}) consider two
distributions $g_1$ and $g_2$ which satisfy the following second-order
equations:
\beq
\label{gf:2}
 \left( \Box - \beta^2 \right)g_1 &=& 4\pi \delta^{(4)}\left( x-x'\right) \,, \\
\label{gf:3}
 \Box g_2 &=& 4\pi \delta^{(4)} \left( x-x'\right) \,.
\eeq
Then one can easily verify that the combination 
\be
\label{gf:4}
 G_{\bf R}(x,x') = \frac{1}{\beta^2}\left( g_1 - g_2\right) \,, \ee
will be a solution to Eq.~(\ref{gf:1}).
Physically we are interested in the retarded Green's function 
solution to Eq.~(\ref{gf:1}), that is of the form
\beq
\label{gf:5}
G_{\rm R}(x,x') = \Theta(t-t') g(x-x')\, ,
\eeq
where $\Theta(z)$ is the Heavyside step function. So $g_1$ and $g_2$
in Eq.~(\ref{gf:4}) above must each be retarded solutions to
Eqs.~(\ref{gf:2}) and~(\ref{gf:3}), respectively. The other three
combinations of retarded and advanced solutions violate causality and
will not be considered further.
So we have reduced the problem of finding the retarded Green's
function solution of the fourth-order, Eq.~(\ref{gf:1}), to finding the
retarded Green's function solutions of the two second-order
differential equations, Eqs.~(\ref{gf:2}) and~(\ref{gf:3}).  The
explicit calculations are given in the Appendix and result in: 
\beq
\label{gf:6}
g_{1\rm R} &=& \left\{\frac{\beta {\cal J}_1(\beta\tau )}{\tau} \Theta(cT -
|\bR|) - 2\delta (\tau^2) \right\}\Theta(T) \, ,\\
\label{gf:7}
g_{2\rm R}  &=&  -2\delta (\tau^2) \Theta(T) \, ,
\eeq
where we have defined
\beq
T&\equiv& t-t'\nonumber\\
\bR &\equiv& \br - \br'\nonumber\\
\tau &\equiv&\sqrt{(cT)^2 - |\bR|^2}\nn~,
\eeq
and ${\cal J}_1(x)$ is the first order Bessel function of the first kind.
Thus, Eq.~(\ref{gf:4}) implies that
\beq
\label{gf:8}
G_{\rm R}(x,x') &=& \frac{ {\cal J}_1(\beta\tau )}{\beta\tau} \Theta(cT -
|\bR|)\Theta(T) \, .  \eeq 
Note the absence of the delta function singularities on the light
cone, consistent with the general analysis detailed in
Ref.~\cite{Prakash:1950}.

Using Eq.~(\ref{eq:field}) one finds that the field is given by
\beq h^{\mu\nu}\left(\br,t\right) =\frac{4G\beta}{c^4} \int {\rm
  d}\br' {\rm d} t'\frac{\Theta\left(T\right)}{\sqrt{\left(cT\right)^2
    - |\bR|^2}}\nonumber\\ \times {\cal J}_1\left(
\beta\sqrt{\left(cT\right)^2 - |\bR|^2}\right) T^{\mu\nu}\left(
\br',t'\right)\nonumber\\ \times \Theta\left( cT - |\bR|\right) ~.
\eeq
One thus finds that the field is sourced only by regions {\it within}
our past light code ({\sl i.e.}, $cT > |\bR|$ ), which is expected for
the propagation of a (positive-real) massive field.  Notice that if
$\beta^2 <0$, corresponding to $\alpha_0>0$, we find that an observed
field is sourced from regions with space-like separation. This is due
to the fact that $\beta^2<0$ corresponds to a tachyon (complex mass)
mode of the gravitational wave.

If we consider the far-field limit, {\sl i.e.} $|\br| \approx |\br -
\br'|$, we can write the spatial components of this field as
\beq\label{eq:4} h^{ik}\left( \br,t\right) \approx \frac{2G
  \beta}{3c^4} \int_{-\infty}^{t-\frac{1}{c}|\br|} \frac{{\rm
    d}t'}{\sqrt{c^2\left( t-t'\right)^2 - |\br|^2} }\nonumber\\
\times {\cal J}_1 \left(
\beta\sqrt{c^2\left( t-t'\right)^2 - |\br|^2}\right)
\ddot{D}^{ik}\left(t'\right)~, \eeq
where we have, as usual, introduced the quadrupole moment,
\be D^{ik}\left(t\right) \equiv \frac{3}{c^2}\int {\rm d}\br \ 
x^i x^k T^{00}(\br,t)~, \ee
the second time derivative of which is given by
\be \int {\rm d}\br T^{ik}\left( \br,t\right) = -\frac{1}{6}\frac{{\rm
    d}^2  }{{\rm d} t^2}\Big[D^{ik}(t)\Big]~.  \ee

In conclusion, just as in the case of General Relativity, we find that
the gravitational waves are only sourced from systems with a
nontrivial quadrupole moment. This is essentially due to the
conservation of the energy momentum tensor, which is unaltered in this
theory. However, the propagation of gravitational waves is
significantly altered, by the presence of additional massive modes.

\section{Examples}
\label{examples}
As a simple pedagogical example let us consider the case of
$\ddot{D}^{ik} \approx {\rm constant}$, for which one can explicitly
perform the integral in Eq.~(\ref{eq:4}) to find
\beq
 h^{ik}\left( \br, t\right) &\approx& -\frac{2G}{3c^2|\br|}
\ddot{D}^{ik}\Big|_{\rm fixed}\nonumber\\
&&\times \left[ \sinh \left( \beta|\br|\right) -
\cosh\left( \beta|\br|\right)+1\right]~.  
\eeq
Thus, one recovers the standard result in the $\beta\rightarrow
\infty$ limit, namely
\beq \lim_{\beta\rightarrow \infty} h^{ik}\left( \br,t\right)
&=& \lim_{\beta\rightarrow\infty}\Big[
-\frac{2G}{3c^4|\br|}\ddot{D}^{ik} \Big|_{\rm fixed}\left( 1 -
e^{-\beta|\br|}\right)\Big]\nonumber \\
&=& ^{({\rm
    GR})}h^{ik}\left(\br, t\right)\lim_{\beta\rightarrow\infty}\left(
1 - e^{-\beta|\br|}\right), \eeq
where $^{({\rm GR})}h^{ik}$ denotes the field in the General
Relativistic case. This is expected, since the $\beta \rightarrow
\infty$ limit corresponds to taking $\alpha_0\rightarrow 0$ and, as it
can be seen from either Eq.~(\ref{eq:0}) or Eq.~(\ref{eq:1}), one then
recovers the Einstein-Hilbert result.

The next simplest example is a system with a periodically varying
quadrupole moment, {\sl i.e.}
\be \ddot{D}^{ij}\left( t\right) = A^{ij}
\cos\left( \omega^{ij} t +\phi^{ij}\right)~, \ee 
where $A^{ij}$ is a constant in time, $\omega^{ij}$ is the frequency
of the oscillations of the $ij$ component and $\phi^{ij}$ is the phase,
both of which we consider to be
time independent. No summation over $i$,
$j$ is implied. Using this in Eq.~(\ref{eq:4}) one finds
\beq\label{eq:grav_pert} \dot{h}^{ij} &=& \frac{4G\beta
  A^{ij}\omega^{ij}}{3c^4} \Biggl[ \sin\left( \omega^{ij} t +
  \phi^{ij} \right) f_c\left( \beta |{\bf r}|,
  \frac{\omega^{ij}}{\beta c}\right) \nonumber \\ &&+ \cos\left(
  \omega^{ij}t+\phi^{ij}\right) f_s\left( \beta |{\bf r}|,
  \frac{\omega^{ij}}{\beta c}\right) \Biggr]~, \eeq
where again no summation is implied and we have defined the functions,
\beq\label{eq:f1}
 f_{\rm s}\left( x,z\right) &\equiv& \int_0^\infty
\frac{{\rm d}s}{\sqrt{s^2 + x^2}} {\cal J}_1\left(s\right) \sin
\left(z\sqrt{ s^2 + x^2} \right)~,\nonumber\\
\label{eq:f2}
f_{\rm c}\left( x,z\right) &\equiv&
\int_0^\infty \frac{{\rm d}s}{\sqrt{s^2 + x^2}} {\cal
  J}_1\left(s\right) \cos \left(z\sqrt{ s^2 + x^2} \right).\nonumber \\
\eeq
These functions are highly oscillatory, with somewhat different
behavior for $z>1$ and $z<1$. Because they have a typical frequency of
the order of $z$, which is (in general) different to $\omega^{ij}$,
the wave-form of the gravitational radiation, Eq.~(\ref{eq:4}) and its
time derivative, Eq.~(\ref{eq:grav_pert}) can experience beat
phenomena. In particular, interference between the various functions
can result in a significant enhancement of the amplitude.

As a specific example, of significant physical interest, consider a
pair of masses $m_1$ and $m_2$, in a circular binary system, under the
assumption that the internal structure of the bodies can be
neglected. For such a system, orbiting in the $xy$-plane, one finds
that the only nonzero components of the quadrupole
are~\cite{landau_lifshitz},
\beq \ddot{D}^{xx}\left(t\right) &=& 12 \mu |\rho|^2 \sin \left(
2\psi\left(t\right)\right)\omega^3 \nonumber\\ &=&
-\ddot{D}^{yy}\left(t\right)~, \nonumber
\\ \ddot{D}^{xy}\left(t\right) &=& -12 \mu |\rho|^2 \cos\left(
2\psi\left(t\right) \right)\omega^3~,\nonumber \\ D^{zz} &=& - \mu
|\rho|^2~, \eeq
where $\mu = m_1m_2/(m_1+m_2)$ is the reduced mass of the system,
$|\rho|$ is the magnitude of the separation vector between the bodies,
which is constant for circular orbits, $\psi$ is the angle of the
bodies relative to the $x$-axis and $\omega=\dot{\psi}$ is the orbital
frequency, which for this simple system is a constant given by
\be \omega \equiv \dot{\psi} = |\rho|^{-3/2} \sqrt{ G\left( m_1 +
  m_2\right)}~.  \ee
Using Eq.~(\ref{eq:4}) one finds
\beq
&&\dot{h}^{ij}\dot{h}_{ij}= \frac{9\mu^2|\rho|^2 \omega^4 G^2
  \beta^2}{c^6}\nonumber\\
&&~~~~~~~~ \times \left[ f_{\rm c}^2\left(\beta|{\bf
    r}|,\frac{2\omega}{\beta c}\right) + f_{\rm s}^2\left(\beta|{\bf
    r}|,\frac{2\omega}{\beta c}\right)\right]~. \eeq
Following the standard approach ({\sl see e.g.},
Ref.~\cite{landau_lifshitz}), one finds that the rate of energy loss
from a system, in the far field limit, is given by
\be\label{eq:energy} -\frac{{\rm d} {\cal E}}{{\rm d}t} \approx
\frac{c^2}{20G} |{\bf r}|^2 \dot{h}_{ij} \dot{h}^{ij}~.  \ee
This allows us to explicitly test the approximation against binary
pulsar measurements, for which the energy loss has been very well
characterized~\cite{c_will}.

For a quantitative fit to the data one would have to extend this
example to non-circular orbits and also account for tidal and other
near field effects. However, such calculations rapidly become rather
involved and even within this simple (and important) system one can
derive some general consequences.

\begin{figure}
 \begin{center}
  \includegraphics[scale=0.8]{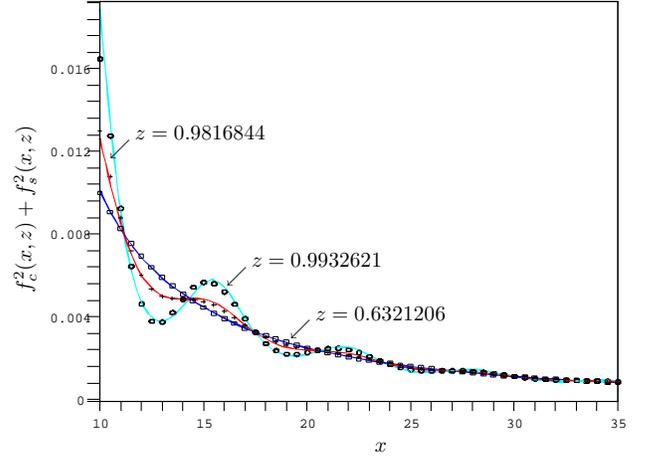}
  \caption{ \label{fig:2} The points are the numerical evaluation of
    $f_{\rm c}^2(x,z) + f_{\rm s}^2(x,z)$ (for three different values
    of $z<1$) and the lines are plots of the corresponding fitted
    function given in Eq.~(\ref{eq:fit}). Notice that the fitting
    function breaks down as we approach $z\rightarrow 1$, which
    corresponds to $2\omega \rightarrow \beta c$.}
 \end{center}
\end{figure}

The functions in Eq.~(\ref{eq:f1}) are highly resonant at $z=1$, which
corresponds to an orbital frequency $\omega = \beta c/2$, however they
are readily calculated for both $z<1$ and $z>1$.  In these regions the
functions given in Eq.~(\ref{eq:f1})  can be evaluated numerically and
fitted to an explicit functional forms. For $\omega < \beta c/ 2$ this
gives
\beq\label{eq:fit} \Big[f_{\rm c}\left(\beta|{\bf
    r}|,\frac{2\omega}{\beta c}\right)\Big]^2 + \Big[f_{\rm
    s}\left(\beta|{\bf r}|,\frac{2\omega}{\beta c}\right)\Big]^2
\nonumber\\ \approx \frac{1}{\left( \beta |{\bf r}|\right)^2}
\exp\left( \frac{C}{\beta |{\bf r}| \left( 1-
  \frac{2\omega}{c\beta}\right) } {\cal J}_1\left( \beta |{\bf r}| -
\frac{2\omega}{c\beta}\right)\right)~, \eeq
where $C\approx 0.175$ is approximately a constant except as
$2\omega \rightarrow\beta c$. In Fig.~\ref{fig:2} we illustrate some
examples of this approximation and show that the approximation is
good even for $2\omega/\beta c \approx 0.99$.

\begin{figure}
 \begin{center}
 \includegraphics[scale=0.7]{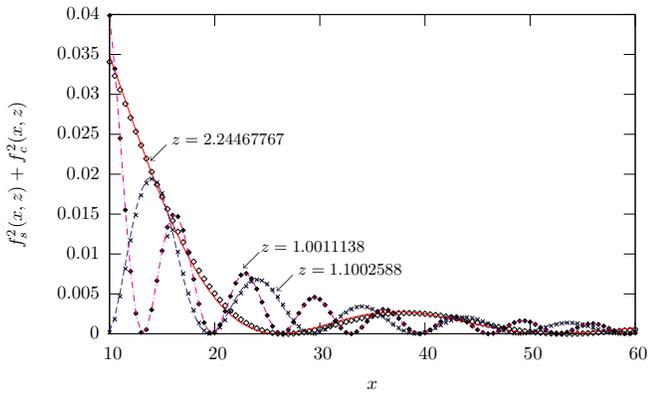}
  \caption{ \label{fig:3} The points are the numerical evaluation of
    $f_{\rm c}^2(x,z) + f_{\rm s}^2(x,z)$ (for three different values
    of $z>1$) and the lines are plots of the corresponding fitted
    function given in Eq.~(\ref{eq:fit2}). Notice that the fitting
    function remain a good approximation even as $z\rightarrow 1$, which
    corresponds to $2\omega \rightarrow \beta c$.}
 \end{center}
\end{figure}

Whilst for $\omega > \beta c/2$ one finds
\beq\label{eq:fit2}
&&\Big[f_{\rm c}\left(\beta|{\bf r}|,\frac{2\omega}{\beta
    c}\right)\Big]^2 + \Big[f_{\rm s}\left(\beta|{\bf
    r}|,\frac{2\omega}{\beta c}\right)\Big]^2 \nonumber \\\approx &&
\frac{4}{\left( \beta |{\bf r}|\right)^2 } \sin^2\left( \beta |{\bf
  r}| \left( \tilde{f}\left(\frac{2\omega}{\beta c}\right)\right)^{-1}
\right)~, \eeq
where for the function $\tilde{f}$ is approximately
\beq \tilde{f}\left( \frac{2\omega}{\beta c}\right)&\approx&
 4\sqrt{
  \left(\frac{2\omega}{\beta c}\right)^2 - 1 } \nonumber\\
&& + 2\exp \left( -\sqrt{
  \left(\frac{2\omega}{\beta c}\right)^2 -1}\right)~.  \eeq
Figure~\ref{fig:3} shows some examples of the quality of this fit,
which, just as for Eq.~(\ref{eq:fit}), is best away from $\omega =
\beta c/2$, however remains very good even as one approaches this
limit. It is important to note that Eq.~(\ref{eq:fit2}) does not
approximate the General Relativity solution for $\beta \rightarrow
\infty$, since, in this limit, all (finite) orbital frequencies
satisfy $\omega < \beta c/2$.

Using the approximation given in Eq.~(\ref{eq:fit}), one can check
that, for slow orbital frequencies, the expected result of General
Relativity is indeed recovered in the $\beta \rightarrow \infty$
limit.  Specifically, we can expand Eq.~(\ref{eq:energy}) in the large
distance (large $|{\bf r}|$) and small orbital frequency ({\sl i.e.},
$2\omega \ll c\beta$) limit, to find the first order corrections to
the standard result of General Relativity, namely
\beq\label{eq:GR_approx} -\frac{{\rm d} {\cal E}}{{\rm d}t} \approx
\frac{32 G \mu^2 \rho^2 \omega^6}{5c^5}
\ \ \ \ \ \ \ \ \ \ \ \ \ \ \ \ \ \ \ \ \ \ \ \ \ \ \ \ \ \ \ \ \ \ \ \ \ \ \ \ \nonumber
\\ \times\left[ 1 + \frac{C}{\beta|{\bf r}| \left( 1-
    \frac{2\omega}{\beta c}\right)} {\cal J}_1 \left( \beta |{\bf r}|
  - \frac{2\omega}{\beta c}\right) +\dots \right]~.  \eeq
Thus, the $\beta \rightarrow \infty$ ({\sl i.e.}, $\alpha_0
\rightarrow 0$) limit reproduces the General Relativity result, as it
should. Also note that any deviation from the standard result is
suppressed by the distance to the source, at least for orbital
frequencies small compared to $\beta c$. Although the amplitude of the
deviation from the standard result is small, there are two interesting
features: firstly, the existence of a maximum frequency $\beta c$ and
secondly, the fact that the deviation is oscillatory.

The maximum frequency comes from the fact that, in addition to a
natural speed $c$, this theory has a natural length, given by
$\beta^{-1} = \sqrt{-\alpha_0 G}$. This natural length scale comes
from the first two moments of the test function used to define the
spectral action, Eq.~(\ref{eq:moments}). Physically, one can think of
this as the scale at which noncommutative effects become
dominant. This is extremely suggestive of an underlying maximum
frequency, which would have rather significant consequences for the
particle physics sector of the theory, in particular for
renormalization. However, it must be remembered that here we are
working with one simple system and even in this case
Eqs.~(\ref{eq:fit})-(\ref{eq:fit2}) are numerical
approximations. Nonetheless, the existence of a maximum frequency in
this system allows NCG effects to significantly enhance the production
of gravitational radiation. This is particularly important given the
suppression with $1/|{\bf r}|$, that is present in
Eq.~(\ref{eq:GR_approx}).

The presence of the Bessel function in Eq.~(\ref{eq:GR_approx}) means
that the amplitude of the deviation from the standard result of
General Relativity will oscillate both with changing distances and
changing frequencies. This allows for a myriad of possible
observational signatures, such as distinct beats of the observed
energy loss of binary pulsars, correlated to their changing orbital
frequency and the distance to the binary. The fact that a similar
phenomena occurs for the gravitational wave itself, Eq.~(\ref{eq:4}),
implies that there would also be a beat structure in direct detection
observations. In the case of a binary pair the amplitude of the beat
will be heavily suppressed compared to that of the carrier wave and
thus it is likely to remain below observational sensitivity,
except in systems with very large orbital frequency.

\section{Conclusions}
\label{conclusions}
NonCommutative Geometry is a natural extension to our familiar notions
of Riemannian geometry, that has the additional benefit of producing
the action of all the Standard Model fields in addition to gravity
terms, purely through geometrical considerations. Thus NCG treats both
gravity and matter on an equal footing and provides us with concrete
relationships between matter and gravitational couplings. The
gravitational sector of (the asymptotic expansion of) this theory
produces modifications to General Relativity and in this paper we
explore the ramifications of these modifications on the formation and
evolution of gravitational waves.

We have shown that the theory contains both massive and massless
gravitons and that the requirement that the mass of these gravitons be
positive fixes the sign of one of the couplings in the theory (for a
given choice of sign conventions). We also show that both these
modes are sourced by the quadrupole moment of a system (just as in
standard GR) and that the retarded Green's function is not restricted
to the past light cone of the observer (unlike GR), as one would
expect for a system with massive modes. We have explicitly calculated
the energy loss for a circular binary system and compared the results
to those of standard GR. We have demonstrated that the amplitude of
these NCG modifications is suppressed by the distance between the
observer and the source of the gravitational waves and hence will
typically be small.

Despite the extremely small amplitude of deviations from standard
results, we have shown that NCG produces several distinctive
features. Firstly, the amplitude of the energy lost by a binary pair
can be higher or lower than the expected value, depending on the
orbital period of the pair and the distance to the observer. This
opens up the possibility that the observed energy loss from such a
pair would be seen to oscillate as the binary moves with respect to
the Earth. Whilst such effects are likely to be beyond current
observational resolution, they allow for an unexpected beat
phenomenon, which would be a concrete signature of NCG.

In addition, we have shown that the amplitude of these effects is
(approximately) proportional to $(1-2\omega/c\beta)^{-1}$, where
$\omega$ is the orbital frequency of the binary. Thus, it would appear
that the NCG corrections to the energy loss by the binary can become
arbitrarily large as the frequency of the binary approach the critical
frequency $\beta c$. In such a regime, the weak field approach taken
here would no longer be valid (and numerical approximations break
down), so one would not trust systems very close to this limit,
however it is certainly true that astrophysical constraints on the
parameters of the theory will be
significantly improved for objects with a very rapidly changing
quadrupole moment. A precise understanding of such systems is likely to
require detailed knowledge of various astrophysical effects (radiation
and particle production, tidal stripping etc.) as well as analytic
solutions to the graviton field in the large frequency regime.

Finally, the form of Eq.~(\ref{eq:grav_pert}) suggests that similar
behaviour may be present in other systems, with periodic, or almost
periodic, variations in the (mass) quadrupole moment.  For laboratory
systems, the gravitational radiation predicted by General Relativity
is negligible, however if the NCG enhancement were sufficiently large,
this may no longer be true. Laboratory systems regularly have very
large oscillation frequencies ({\rm e.g.}, lattice vibrations in
solids can easily exceed $10^{12} {\rm Hz}$) which would experience
anomalous damping, if the system was producing significant amounts of
gravitational radiation. This opens up the (remote) possibility that
the noncommutative nature of space-time might be probed in the
laboratory.

One can immediately use the results of this paper to examine circular
binary systems, in order to constrain the value of
$\beta$~\cite{wjm}. Similarly, one can include eccentricity which may
result in more restrictive constraints on the theory. An alternative
avenue would be to use the gravitational wave-forms given here to
deduce the consequences for direct gravity wave searches (LIGO, VIRGO,
LISA, {\sl etc}).  In particular, to extend these result to the large
field regime and look for modifications to the {\sl chirp} that
develops at the end of in-spiral events.

\vskip1truecm
{\bf Acknoweldgments}
\medskip
\\
The work of W.\ N.\ is supported in part by the NSF grant PHY0854743,
the George A. and Margaret M. Downsbrough Endowment and the Eberly
research funds of Penn State.  The work of M.\ S.\ is partially
supported by the European Union through the Marie Curie Research and
Training Network {\sl UniverseNet} (MRTN-CT-2006-035863). J.\ O.\ 
acknowledges support from the Alfred P. Sloan Foundation and the 
Eberly College of Science.




\setcounter{equation}{0}  
\renewcommand{\theequation}{A-\arabic{equation}}
\section*{Appendix}
In what follows, we shall explicitly detail the calculations which
were performed in order to determine the Green's functions, namely
Eqs.~(\ref{gf:6}), (\ref{gf:7}).

Consider first the following fourth-order wave equation
\beq
\label{A:1}
\Big (\Box-\beta^2\Big)\Box G_{\rm R}(x,x') \,  &=& 4\pi\delta^{(4)}(x-x') \, .
\eeq
Lorentz symmetry of the background restricts $G_{\rm R}$ to be solely a 
function of $x-x'$. Given the following inverse Fourier transforms:
\be
\label{A:2}
 G_{\rm R}(x,x') = \frac{1}{(2\pi)^4}\int {\rm d}^4k\, \tilde G_{\rm
   R}(k)e^{ik\cdot (x-x')}\, , \ee
and
\be
\label{A:3}
 \delta(x-x') = \frac{1}{(2\pi)^4} \int {\rm d}^4k\, e^{ik\cdot
   (x-x')} \, , \ee
where $k\cdot z = -\omega z_0 +{\bf k} \cdot {\bf z}$, the Fourier transform 
$\tilde G_{\rm R}(k)$ must satisfy
\beq
\label{A:4}
\tilde G_{\rm R}(k) = \frac{4\pi}{\big[(\omega+i\epsilon)^2 - {\bf k}^2 -\beta^2
\big ]\big [(\omega+i\epsilon)^2 -{\bf k}^2 \big ]} \, ,
\eeq
in order to solve Eq.~(\ref{A:1}).  

\begin{figure}
\begin{center}
\includegraphics[scale=0.6]{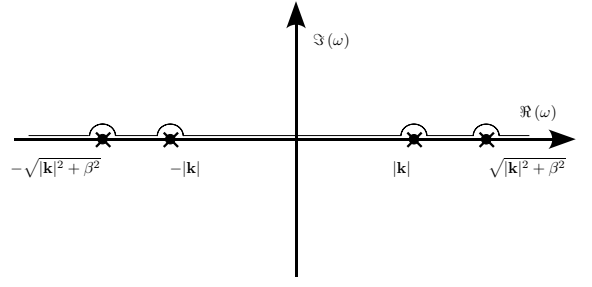}
\caption{\label{Fig:A1} The retarded Green's function is defined by
  extending the contour of the integral into the positive half of the
  imaginary plane around each of the four poles, $\omega = \pm |{\bf
    k}|$ and $\omega = \pm \sqrt{|{\bf k}|^2 + \beta^2}$.}
\end{center}
\end{figure}

Upon performing the inverse Fourier transform to determine the
coordinate expression for $G_{\rm R}(x,x')$, the following pole
prescription uniquely determines the retarded Green's function (the
reader is referred to Fig.~\ref{Fig:A1}). The contours are traversed
above the poles in the complex $\omega$ plane: for $t-t'>0$ we close
the contour in the lower half plane picking up the residue of the
poles; for $t-t'<0$ we close the contour in the upper half plane, thus
enclosing no poles.

Note that $\tilde G_{\rm R}(k)$ can be rewritten as 
\beq
\tilde G_{\rm R}(k) &=& 4\pi \left\{
\frac{1}{\beta^2\big[(\omega+i\epsilon)^2 - {\bf k}^2 -\beta^2 \big
]}\right.\nonumber\\ \nn &&\left.\qquad\qquad-\frac{1}{\beta^2 \big
  [(\omega+i\epsilon)^2 -{\bf k}^2 \big ]}\right\}\, ,\nonumber\\
 \label{A:6}
&\equiv& \frac{1}{\beta^2} \Big[ \tilde g_1(k) - \tilde
 g_2(k)\Big] \, , \eeq
which is in fact simply the Fourier transform of Eq.~(\ref{gf:4}). We
will analyze each term in Eq.~(\ref{A:6}) separately. 

First we need to define the following quantities:
\beq
T&\equiv& t-t'\nonumber \, ,\\
\bR &\equiv& \br - \br'\nonumber \, ,\\
\tilde{k}&\equiv& |{\bf k}|^2 + \beta^2 \, .
\eeq
In what follows, we will set the speed of light $c=1$. The $\tilde
g_1(k)$ term can be integrated as follows: 
\beq
\label{A:7}
g_{1 {\rm R}}(x-x') &=& \frac{1}{(2\pi)^4}\int {\rm d}\omega {\rm
  d}^3{\bf k}\, \tilde g_1(k)\nonumber\\ \nn &=& \frac{1}{4\pi^3}\int {\rm
  d}\omega {\rm d}^3{\bf k}\, \nonumber\\
&&\hspace*{-1.4truecm}\times\frac{e^{-i\omega T} e^{i{\bf k}\cdot \bf
    \bR }}{\big[((\omega+i\epsilon) +\sqrt{\tilde{k}})((\omega+i
    \epsilon) -\sqrt{\tilde{k}})\big]} \, , \eeq
where the $+i\epsilon$ above is simply a mnemonic for the retarded
Green's function pole prescription. The contour integral in the
complex $\omega$ plane results in $-2\pi i \sum Res$ and we have
\beq
\label{A:8}
g_{1 {\rm R}}(x-x') =  \frac{
  i\Theta(T)}{4\pi^2}
&&
\nonumber\\
 &&\hspace*{-2.5truecm}\times \int {\rm d}^3{\bf k}\, e^{i{\bf k}\cdot \bR }
\left \{ \frac{e^{i T\sqrt{\tilde{k}}}-e^{-i T
    \sqrt{\tilde{k}}}}{\sqrt{\tilde{k}}}\right \} \, .  \eeq
Upon performing the angular integral, one finds
\be
\label{A:9}
g_{1 {\rm R}}(x-x') = \frac{ \Theta(T)\beta}{2\pi |\bR|}\left[ 
{\cal I}^+ - {\cal I}^- \right]~,
\ee
where
\beq {\cal I}^{\pm} = \int_0^\infty \frac{k{\rm d}k}{\sqrt{k^2+1}}
\Big( e^{i\beta(|\bR|k\pm T\sqrt{k^2 + 1})}\nonumber\\
~~~~~~+e^{-i\beta(|\bR|k \pm
  T\sqrt{k^2 + 1})}\Big)~. \eeq
We will first focus on the solution which is interior to the light
cone, {\sl i.e.}, $T> |\bR|$. In the first term in the integral, 
Eq.~(\ref{A:9}) above, we perform the following change of variables:
\beq \nn \nn  |\bR|k+ T\sqrt{k^2 + 1} &=& \tau \cosh v \, , \\ \nn \nn
 Tk+ |\bR|\sqrt{k^2 + 1} &=& \tau \sinh v \, , \\ \nn \nn  k
&=&-\frac{|\bR|}{\tau}\cosh v +\frac{T}{\tau}\sinh v \, ,\\ \nn \nn 
\frac{{\rm d}k}{\sqrt{k^2 + 1}} &=& {\rm d} v \, , \eeq
where $v\in [v_0,\infty)$, $v_0 \equiv \text{arcosh}\left( T/\tau
  \right )$ and we have defined $\tau =\sqrt{T^2-|\bR|^2}$. 

In the second term in the integral in Eq.~(\ref{A:9}) above, we will
perform a different change of variables given by
\beq \nn \nn |\bR|k- T\sqrt{k^2 + 1} &=& -\tau \cosh \bar v \, ,
\\ \nn \nn  Tk - |\bR|\sqrt{k^2 + 1} &=& \tau \sinh \bar v \, ,\\ \nn
\nn k &=&\frac{|\bR|}{\tau}\cosh \bar v +\frac{T}{\tau}\sinh \bar v
\, .\\ \nn \nn  \frac{{\rm d}k}{\sqrt{k^2 + 1}} &=& {\rm d} \bar v \,
, \eeq
where for these change of variables $\bar v\in [-v_0, \infty)$. Note
  that, the variable $\bar v$ spans the interval $ [-v_0,0]$, while $k$
  correspondingly spans the interval $[0,|\bR|/\tau ]$. After
  these changes of variables we arrive at the following:
\begin{widetext}
\beq
 \label{A:10}
g_{1 {\rm R}}(x-x') &=&\frac{ \Theta(T)\beta}{2\pi |\bR|} \bigg \{
\int_{v_0}^\infty {\rm d}v \Big[-\frac{|\bR|}{\tau}\cosh v
+\frac{T}{\tau}\sinh v \Big ]\Big[ e^{i\beta \tau\cosh v}
+e^{-i\beta\tau\cosh v}\Big ]\nonumber \\ \nn \nn &&\quad \quad
\qquad\,\,\, -\int_{-v_0}^\infty {\rm d}\bar v\Big[\frac{|\bR|}{\tau}
\cosh \bar v +\frac{T}{\tau}\sinh \bar v\Big]\Big[ e^{ i\beta \tau\cosh
  \bar v}+e^{-i\beta \tau\cosh \bar v} \Big] \bigg \} \, , \nonumber\\
 &=&\frac{ \Theta(T)\beta}{2\pi |\bR|} \int_{-\infty}^\infty {\rm d}v
\Big[-\frac{|\bR|}{\tau} \cosh v +\frac{T}{\tau}\sinh v \Big]\Big[ e^{i\beta
  \tau\cosh v}+e^{-i\beta\tau\cosh v}\Big] \,  \nonumber\\
&=& -\frac{ \Theta(T)\beta}{2\pi \tau} \int_{-\infty}^\infty {\rm d}v
\Big[ e^{i\beta \tau\cosh v - v}+e^{-i\beta\tau\cosh v -v} \Big]\, ~, \eeq
\end{widetext} 
where the last equality follows by symmetry; it should be understood
that the limits in the integrals above are such that the integrals are
convergent.

We note the following integral
representations~\cite{Abramowitz&Stegun} of the Hankel functions of
order $\alpha$ of the first and second kind, respectively:
\beq H^{(1)}_\alpha (x) &=&\frac{ e^{-i\alpha\frac{\pi}{2}}}{\pi i}
\int_{-\infty -i\epsilon}^{\infty+i\epsilon} {\rm d}v\, e^{ix\cosh v -
  \alpha v}~,\nonumber\\ H^{(2)}_\alpha (x) &=& -\frac{
  e^{+i\alpha\frac{\pi}{2}}}{\pi i}
\int_{-\infty+i\epsilon}^{\infty-i\epsilon} {\rm d}v\, e^{-ix\cosh v -
  \alpha v}~, \nonumber \eeq
related to the Bessel functions of first and second kind via the
relations:
\beq
\label{A:15}
 H^{(1)}_\alpha (x) = {\cal J}_\alpha (x) +i{\cal Y}_\alpha (x) \, ,  \\
 \label{A:16}
 H^{(2)}_\alpha (x) = {\cal J}_\alpha (x) -i{\cal Y}_\alpha (x) \, ,
\eeq
where ${\cal J}_\alpha (x)$ and ${\cal Y}_\alpha (x)$ are Bessel functions of the
first and second kind respectively; {\sl i.e.}, the two linearly
independent solutions to Bessel's equation:
\be
\label{A:17}
 x^2 \frac{d^2F}{dx^2} + x \frac{dF}{dx} +(x^2 - \alpha^2) F = 0 \, .
\ee
In particular, we require the Hankel functions of order $\alpha = 1$,
which are given as 
\beq \nn H^{(1)}_1 (x) \nn &=&-\frac{1}{\pi }
\int_{-\infty -i\epsilon}^{\infty+i\epsilon} {\rm d}v\, e^{ix\cosh v -
  v}\, , \\ \nn H^{(2)}_1 (x) \nn &=& -\frac{ 1}{\pi }
\int_{-\infty+i\epsilon}^{\infty-i\epsilon} {\rm d}v\, e^{-ix\cosh v -
  v} \, .  \eeq 
Thus, one can express the integral for $g_{1 {\rm R}}$,
Eq.~(\ref{A:10}), in terms of these Hankel functions as 
\beq\label{A:18} \nn g_{1 {\rm R}}(x-x') &=&\frac{
  \Theta(T)\beta}{2\pi \tau} \left \{ \pi H^{(1)}_1(\beta\tau ) +\pi
H^{(2)}_1(\beta\tau ) \right \} \, , \nonumber\\ &=&
\Theta(T)\frac{\beta {\cal J}_1(\beta\tau )}{\tau} \, \,\quad\mbox{for}\ \ T>
|\bR|\, . \eeq
Note that, ${\cal J}_1(\beta\tau )$ is the Bessel function of first kind of
order $1$ and we have used Eqs.~(\ref{A:15}) and (\ref{A:16}) to
arrive at the final expression, Eq.~(\ref{A:18}), above.

Looking now at the exterior of the light cone, {\sl i.e.}, $|\bR| >
T$, we make the following change in variables in the first term in the
integral of Eq.~(\ref{A:9}) above:
\beq \nn \nn  |\bR||{\bf k}|+
T\sqrt{|{\bf k}|^2 + 1} &=& \xi \sinh v \, , \\ \nn \nn T|{\bf k}|+
|\bR|\sqrt{|{\bf k}|^2 + 1} &=& \xi \cosh v \, ,\\ \nn \nn  k
&=&\frac{|\bR|}{\xi}\sinh v -\frac{T}{\xi}\cosh v \, ,\\  
\nn \nn \frac{{\rm d}|{\bf k}|}{\sqrt{|{\bf k}|^2 + 1}} &=& {\rm d} v \, .  \eeq
where as before $v\in [v_0,\infty)$, but now $v_0 \equiv
  \text{arsinh}\left( T/\xi \right )$ and we have defined $\xi
  =\sqrt{|\bR|^2- T}$. 

In the second term in the integral of Eq.~(\ref{A:9}) we perform the
following change of variables:
\beq
 \nn \nn   |\bR||{\bf k}|- T\sqrt{|{\bf k}|^2 + 1} &=& \xi \sinh \bar
  v \, , \\  \nn \nn  -T|{\bf k}|+ |\bR|\sqrt{|{\bf k}|^2 + 1} &=& \xi
  \cosh \bar v \, ,\\  \nn \nn k &=&\frac{|\bR|}{\xi}\sinh \bar v
  +\frac{T}{\xi}\cosh \bar v \, ,\\ \nn \nn  \frac{{\rm d}|{\bf
      k}|}{\sqrt{|{\bf k}|^2 + 1}} &=& {\rm d} \bar v \, , \eeq 
where
  $\bar v\in [-v_0,\infty)$. 

Upon using these change of variables we arrive at 
\beq
\label{A:19}
 g_{1 {\rm R}}(x-x') &=& -2\frac{ \Theta(T)}{ |\bR|} \frac{T}{\xi}
 \delta ( \sqrt{|\bR|^2- T^2}\,)\, ,  \eeq
which vanishes since we are explicitly considering the $|\bR| > T $
region and thus the delta function vanishes. 

We still have yet to determine the singular part of the Green's
function on the light cone $|\bR| = T $. To do this we will repeat the
formalism established in Ref.~\cite{Poisson:2004lr}. Note that, we
very well could have determined the full Green's function $g_{1 {\rm
    R}}(x-x')$ and not just the part on the light cone via the
formalism in Ref.~\cite{Poisson:2004lr}.  In fact, a full review of
the formalism will serve as a useful check on the calculation of the
smooth part of the Green's function determined above via Fourier
transform.

To begin, we first integrate the Green's function equation,
Eq.~(\ref{A:1}), over a space-time volume which contains the source
event $x'=0$, namely
\beq
\label{A:20}
 \int_{\p V}\, (g_1)^{;\mu}{\rm d}\Sigma_\mu  -\beta^2\int_V g_1 =4\pi \, ,
\eeq
where Gauss' theorem was used to arrive at the first term above and
${\rm d}\Sigma_\mu$ is the surface element of the boundary $\p
V$. Assuming $\int_V f_1$ vanishes as the integration volume vanishes,
we are left with 
\beq
\label{A:21}
 \lim_{V\rightarrow 0}\int_{\p V}\, (g_1)^{;\mu}{\rm d}\Sigma_\mu  =4\pi \, .
\eeq
Now introduce the coordinates $(w,\chi, \theta, \phi)$ given by
\beq
t\nn &=&w\cos\chi \, ,\\
\nn x &=& w \sin\chi\sin\theta\cos\phi\, ,\\
\nn y &=& w \sin\chi\sin\theta\sin\phi\, , \\
\nn z &=& w \sin\chi\cos\theta \, ,
\eeq
such that the line element ${\rm d}s^2 = g_{\alpha\beta}{\rm
  d}x^\alpha{\rm d}x^\beta$ of the flat background takes the form
\beq
\label{A:22}
{\rm d}s^2 &=& -\cos2\chi{\rm d}w^2 +2w\sin2\chi{\rm d}w{\rm d}\chi \nonumber\\
&& + w^2\cos2\chi{\rm d}\chi^2 +w^2\sin^2\chi{\rm d}\Omega^2 \, ,
\eeq
where ${\rm d}\Omega^2= {\rm d}\theta^2 +\sin^2\theta {\rm d}\phi^2$.
In these coordinates, the surface $\p V$ is given by constant $w$, and
the Synge world function $\sigma$ is 
\beq
\label{A:23}
 \sigma \nn &=&-\half w^2\cos2\chi~;\\
\eeq
notice that for time-like events $-2\sigma = \tau^2$ where $\tau$ is 
as previously defined.

The following quantities will be useful for what follows:
\beq
\sqrt{-g} \nn &=& w^3\sin^2\chi\sin\theta \, ,\\ g^{ww}\nn
&=&-\cos2\chi \, ,\\ g^{w\chi}\nn &=&\frac{\sin 2\chi}{w} \,
,\\ g^{\chi\chi}\nn &=&\frac{\cos 2\chi}{w^2}\, , \eeq
where $g$ is the determinant of the metric $g_{\alpha\beta}$. The only
nonzero component of ${\rm d}\Sigma_\alpha$ is 
\beq
\label{A:24}
{\rm d}\Sigma_w = w^3\sin^2\chi {\rm d}\chi{\rm d}\Omega \, ,
\eeq
where ${\rm d}\Omega =\sin\theta{\rm d}\theta{\rm d}\phi$. 

In these coordinates, the retarded Green's function is given by:
\beq
\label{A:25}
g_{1\rm R} &=& \Theta( w \cos \chi) g( \sigma )~,
\eeq
where $g(\sigma)$ is an as yet undetermined, possibly distributional, 
function. We will only need the following gradient of $g_{1\rm R}$ 
(omitting the label $\phantom{}_{\rm R}$)
\beq
 (g_1)^{;w}&=&g^{w\mu}(g_1)_{;\mu}\nonumber\\
&=&g^{ww}(g_1)_{;w}+g^{w\chi}(g_1)_{;\chi} \, .
\eeq
A straight-forward calculation leads to 
\beq
\label{A:26}
(g_1)^{;w}  &=& -\delta( w \cos\chi) \cos\chi g(\sigma) \\
\nn && + w\Theta( w \cos\chi)g'(\sigma) \, ,
\eeq
where the prime on $g(\sigma)$ denotes differentiation with 
respect to $\sigma$. We then have
\beq \nn \int_{\p V}\, (g_1)^{;\mu}{\rm d}\Sigma_\mu &=&\int_{\p
  V}w^3\sin^2\chi {\rm d}\chi{\rm d} \Omega\big\{ w\Theta( w
\cos\chi)g'(\sigma) \\ \nn &&\qquad -\delta( w \cos\chi) \cos\chi
g(\sigma)\big\} \,  \\
 \label{A:27}
 &=& 4\pi w^4 \int_0^{\frac{\pi}{2}} \,\sin^2\chi {\rm d}\chi
 g'(\sigma) \, , \eeq
where the Heavyside function has restricted the limits of $\chi$
integration such that $\cos\chi\geq0$ and the delta term vanishes.

Changing integration variable from $\chi$ to $\sigma$ in the integral
above we arrive at the following condition on $g(\sigma)$:
\beq
\label{A:28}
 \lim_{\epsilon\rightarrow 0}\epsilon \int_{-\epsilon}^{\epsilon} {\rm
   d}\sigma \, \Xi \left(\frac{\sigma}{\epsilon}\right) g'(\sigma) = 1
 \, , \eeq
where
\beq
\nn \epsilon &\equiv & \half w^2 \, ,\\
 \nn \Xi \left(\frac{\sigma}{\epsilon}\right) &\equiv & 
\sqrt{\frac{1+\frac{\sigma}{\epsilon}}{1-\frac{\sigma}{\epsilon}}} \, .
\eeq
We now propose the following ansatz for $g(\sigma)$:
\beq
\label{A:29}
g(\sigma) &=& V(\sigma) \Theta(-\sigma) + A\delta(\sigma)
+B\delta'(\sigma)\nonumber \\ && + C\delta''(\sigma) +
D\delta'''(\sigma) +...~,  \eeq
where $V(\sigma)$ is a smooth function of $\sigma$ and $A,B,...$ are
constants.  Inserting the ansatz Eq.~(\ref{A:29}) into
Eq.~(\ref{A:28}) gives
\begin{widetext} 
\beq
 \label{A:31}
 \lim_{\epsilon\rightarrow 0}\epsilon
 \left\{\int_{-\epsilon}^{\epsilon} {\rm d}\sigma \, \Xi
 \left(\frac{\sigma}{\epsilon}\right) V'(\sigma) \Theta(-\sigma)-\Xi
 \left(0\right)V(0) - \frac{A}{\epsilon}\dot\Xi \left(0\right)
 +\frac{B}{\epsilon^2}\ddot\Xi \left(0\right)-
 \frac{C}{\epsilon^3}\dddot\Xi \left(0\right)
 +\frac{D}{\epsilon^4}\Xi^{(4)} \left(0\right) +...\right\} = 1 \, .
 \eeq
\end{widetext} 
The first two terms on the left-hand side of Eq.~(\ref{A:31}) above
vanish, since $V(\sigma)$ is assumed to be smooth. The limit exist as
long as $B=C=...=0$ and the condition Eq.~(\ref{A:28}) will be
satisfied iff $A=-1$, since $\dot\Xi \left(0\right)=1$. To fully
determine the smooth part of $g(\sigma)$ one needs then only to solve
the homogeneous equation ({\sl i.e.}, $x\not = x'$):
\beq \nn (\Box - \beta^2)g(\sigma)=
4g'(\sigma)+2\sigma g''(\sigma)-\beta^2g(\sigma) &=& 0 \, , \eeq
from which it is straight forward to verify that (the reader is referred
to Ref.~\cite{Poisson:2004lr} for further details)
\beq
\label{A:32}
V(\sigma) = \frac{\beta {\cal J}_1(\beta\sqrt{-2\sigma} )}{\sqrt{-2\sigma}} \, .
\eeq
Returning to the original coordinates $(t,r, \theta, \phi)$ we then have
\beq \nn g_{1 {\rm R}}(x-x') \nn &=& - \delta (\half\tau^2) \Theta(T)
\\ \nn&& \,+ \frac{\beta {\cal J}_1(\beta\tau )}{\tau} \Theta\left(\half(T^2
- |\bR|^2)\right)\Theta(T) \, , \\
\label{A:33}
\nn &=& - 2\delta (\tau^2) \Theta(T) \\  &&\, + \frac{\beta
  {\cal J}_1(\beta\tau )}{\tau} \Theta\left(T - |\bR|\right)\Theta(T) \, ,
\eeq
where we have used some standard properties of the Dirac delta function and 
the Heavyside step function. 

To determine $g_{2 {\rm R}}(x-x')$ we need to consider the integral
\beq
\label{A:34}
 g_{2 {\rm R}}(x-x') &=&\frac{1}{4\pi^3}\int {\rm d}\omega {\rm
   d}^3{\bf k}\\ \nn &&\qquad\times\frac{e^{-i\omega T}e^{i{\bf
       k}\cdot \bf \bR }}{\big[(\omega +i\epsilon +|{\bf k}|)(\omega
     +i\epsilon -|{\bf k}|)\big]} \, , \eeq
where the pole prescription is again that of the retarded Green's function. 
A straight-forward calculation leads to
\beq \nn g_{2 {\rm R}}(x-x') &=&\frac{ \Theta(T)}{ |\bR|} \big \{
\delta((T+|\bR|)-\delta( T-|\bR| ) \big\}\, , \\
\label{A:35}
 &=&- 2\Theta(T) \delta( \tau^2) \, .
\eeq
Finally, the Green's function which satisfies the fourth-order wave equation 
Eq.~(\ref{gf:1}) is given by
\beq \nn G_{\rm R}(x-x') &=& \frac{1}{\beta^2}\big(g_{1 {\rm
    R}}(x-x')-g_{2 {\rm R}}(x-x')) \, ,\\
\label{A:36}
 &=&   \frac{\beta {\cal J}_1(\beta\tau )}{\beta\tau}  \Theta(T - |\bR|)\Theta(T)\, .
\eeq
Note that the Green's function above is subject to the initial value
condition:
\be
\frac{\p G_{\rm R}(x-x')}{\p t}\big|_{t=0} = 0~.
\nonumber
\ee

\end{document}